\documentclass{iau} 
\usepackage{graphicx}
\usepackage{xcolor}
\usepackage{url}
\usepackage[colorlinks,urlcolor=teal]{hyperref}

\title[] 
{Long-term optical monitoring \\ of the solar atmosphere in Italy}

\author[S.~L. Guglielmino et al.]   
{S.~L. Guglielmino$^1$, I. Ermolli$^2$, P. Romano$^3$, F. Zuccarello$^1$, \\ F. Giorgi$^2$, M. Falco$^3$, R. Piazzesi$^2$, M. Stangalini$^2$, M. Murabito$^{1,2}$, \\  M. Ferrucci$^2$ \and A. Mangano$^3$}

\affiliation{$^1$Dipartimento di Fisica e Astronomia -- Sezione Astrofisica, Universit\`a di Catania, \\ 
	via S.~Sofia 78, I-95123 Catania, Italy \\ email: {\tt salvo.guglielmino@oact.inaf.it} \\[\affilskip]
	$^2$INAF -- Osservatorio Astronomico di Roma, \\ 
	via Frascati 33, I-00078 Monte Porzio Catone, Italy \\[\affilskip]
	$^3$INAF -- Osservatorio Astrofisico di Catania, \\ 
	via S.~Sofia 78, I-95123 Catania, Italy}

\pubyear{2015}
\volume{xxx}  
\jname{Title of your IAU Symposium}
\editors{A.C. Editor, B.D. Editor \& C.E. Editor, eds.}
\begin{document}

\maketitle

\begin{abstract}
Probably, the long-term monitoring of the solar atmosphere started in Italy with the first telescopic observations of the Sun made by Galileo Galilei in the early $17^{\mathrm{th}}$ century. His recorded observations and science results, as well as the work carried out by other following outstanding Italian astronomers inspired the start of institutional programs of regular solar observations at the Arcetri, Catania, and Rome Observatories.\\ 
These programs have accumulated daily images of the solar photosphere and chromosphere taken at various spectral bands over a time span greater than 80 years. In the last two decades, regular solar observations were continued with digital cameras only at the Catania and Rome Observatories, which are now part of the INAF National Institute for Astrophysics. At the two sites, daily solar images are taken at the photospheric G-band, Blue ($\lambda=409.4$~nm), and Red ($\lambda=606.9$~nm) continua spectral ranges and at the chromospheric Ca~II~K and H$\alpha$ lines, with a $2^{\prime\prime}$ spatial resolution.\\
Solar observation in Italy, which benefits from over 2500 hours of yearly sunshine, currently aims at the operational monitoring of solar activity and long-term variability and at the continuation of the historical series as well. Existing instruments will be soon enriched by the SAMM double channel telescope equipped with magneto-optical filters that will enable the tomography of the solar atmosphere with simultaneous observations at the K~I 769.9~nm and Na~I~D 589.0~nm lines.
In this contribution, we present the available observations and outline their scientific relevance. 

\end{abstract}

\firstsection 
\section{Historical perspective of solar observations in Italy}

The history of long-term solar optical monitoring in Italy dates back to the early $17^{\mathrm{th}}$ century, specifically to the end of 1610 (e.g., \cite[Howard 1998]{Howard98}), when first telescopic observations of the Sun were carried out by Galileo Galilei (1524 -- 1642). His observations of sunspots, which allowed him to publish the treat \textit{Istoria e dimostrazioni intorno alle macchie solari e loro accidenti} (1613),  were recorded in drawings, being now archived in the collections of ancient books of the Arcetri and Rome Observatories. Following Galileo's activity, scientific observations of the Sun flourished at different sites, in particular both in Arcetri (Florence) and in Rome at the ``Collegio Romano''. 

In Rome, solar observations continued during the $17^{\mathrm{th}}$ and $18^{\mathrm{th}}$ centuries with the activity of several outstanding Italian astronomers: Cristoforo Scheiner (1573 -- 1650), whose masterpiece \textit{Rosa Ursina sive Sol} (1630) is worthwhile to be mentioned, Atanasio Kircher (1602 -- 1680), Cristoforo Maire (1697 -- 1767), Ruggero Boscovich (1711 -- 1787) and Giovanni Battista Audifreddi (1714 -- 1794). During the $19^{\mathrm{th}}$ century, the observations carried out at the ``Collegio Romano'' by Ch.~Angelo Secchi, S.J. (1818 -- 1878), Lorenzo Respighi (1824 -- 1889) and Pietro Tacchini (1838 -- 1905) led to the birth of modern solar astrophysics, with their attempts to understand the physics behind the observed phenomena. For example, from observations performed in 1855 with a Merz Telescope at the ``Collegio Romano'', Ch.~Secchi reported on the filamentary and multi-structured nature of sunspot penumbrae.

In this climate, in the late $19^{\mathrm{th}}$ -- early $20^{\mathrm{th}}$ century, solar observations were resumed in Arcetri by Giovan Battista Donati (1826 -- 1873), Gaspare Ferrari (1836 -- 1903) and Giorgio Abetti (1882 -- 1982). Moreover, observations of the Sun began in the sun-drenched island of Sicily, first in Palermo (1869) and then in Catania (1892), with the activity of Pietro Tacchini and Annibale Ricc\`o (1844 -- 1919). Between these different observing locations there were close ties, as witnessed by the moving of Tacchini from Sicily to Rome (1879). In the latter site, Francesco Giacomelli (1849 -- 1936) began solar observations at the ``Campidoglio'' Observatory as well. 

Finally, during the $20^{\mathrm{th}}$ century, synoptic observations of the Sun were carried out in Rome, with an ASKANIA Telescope at ``Campidoglio'' Observatory (1924 -- 1938), which later was moved and operated until 1964 in Villa Mellini -- Monte Mario, where observations were also performed with an Equatorial spar (1947 -- 2001), at the Arcetri Solar Tower (1925 -- 1974), and in Catania (1934 -- 1964). 

Nowadays, solar monitoring in Italy is systematically carried out in two observatories of the INAF -- National Institute for Astrophysics, namely the Astronomical Observatory of Rome (OAR) and the Catania Astrophysical Observatory (OACT).

\section{Overview of historical solar data sets}

The daily work of astronomers who carried out the solar monitoring in Italy over the last centuries resulted in long series of drawings of the photosphere and chromosphere, which now constitute a very valuable collection of historical solar data, mainly archived at the Arcetri, Rome, and Catania Observatories. A large fraction of them is still stored in the original form, i.e., in the collections of ancient books of the above observatories, although a part of the historical data is also available in digital form.

In particular, the OAR archive comprises historical solar observations by Ch.~Secchi and Tacchini at the ``Collegio Romano'' and other astronomers working in Rome. These consist of full-disk drawings of sunspots, faculae, pores, prominences, and ``polar crowns''. Observations by Ch.~Secchi were obtained at the equatorial Merz Telescope (1853, 1858 -- 1896) and Cauchoix telescope (1853, 1855, 1857 -- 1859, 1865 -- 1871). In addition to drawings, the same archive includes tabulated values (position, area) and annotations, among which those of Caterina Scarpellini, a female observer already working at the ``Collegio Romano'' at the time of total solar eclipse of 1860, seen in Rome. 

The OAR archive also includes photographic observations of the solar photosphere and chromosphere acquired at the ASKANIA telescope and stored on plates (1890 -- 1947), and drawings and photographic observations of the photosphere (white light) and chromosphere (H$\alpha$ and Ca~II~K lines) from the monitoring carried out at the Equatorial spar in Monte Mario (1947 -- 2001). Digitazion of the above observations, currently stored on plates and films, is undergoing, with part of them (1964 -- 1970) already digitized.


Moreover, the OAR archive comprises images from the digitization of the photographic observations in Ca~II~K and H$\alpha$ lines acquired at the Arcetri Solar Tower. These cover 5042 days (1926 -- 1974) and were stored on 12759 plates (see \cite[Ermolli et al. 2009]{Ermolli09}). 

The OACT archive contains a century-long ``Sunspot series'' (1865 -- 1964), which consists of 417 archival units comprising drawings of sunspots, faculae and pores, at first observed in Palermo (1862 -- 1891), including notes by Tacchini, then observed in Catania (1892 -- 1964). Note that the period 1926 -- 1933 has been fully digitized. Moreover, there is a ``Prominences series'' (1881 -- 1965), formed by 132 archival units consisting of drawings of prominences at solar limb, including the observations ``Solar limbs as observed at Palermo in 1890'' by Ricc\`o. In addition, there are about 5000 photographic plates of full-disk images taken in the Ca~II~K line at the Equatorial spar (1964 -- 1977).

\section{Overview of current monitoring}


Since 1996, the Precision Solar Photometric Telescope (PSPT) has been operating at OAR (\cite[Ermolli et al. 1998]{Ermolli98}). The PSPT is a low-scattered light, 0.15~m refractor telescope (f=2.3~m) with achromatic doublet designed for synoptic full-disk observations of the photosphere and chromosphere characterized by 0.1\% photometric accuracy. A twin telescope of the PSPT formerly operated at Mauna-Loa by the HAO (1998 -- 2015) . 

The Rome-PSPT is equipped with six narrow-band interference filters and a tip-tilt stabilizing system. At present, it regularly acquires full-disk images in five spectral bands: at the Ca~II~K $393.2$~nm line, with two bandwidths of $0.25$ and $0.11$~nm, at the G band ($430.6 \pm 0.6$~nm), Blue continuum ($409.4 \pm 0.14$~nm), and Red continuum ($606.9 \pm 0.23$~nm), with a spatial resolution of $2^{\prime\prime}$ and a cadence of a few images per day (observing time interval: 8:00 -- 15:00 CET). 

The data are available at the website: \url{http://www.mporzio.astro.it/solare/}.

The Rome-PSPT observations, which are mostly used for studying the photometric properties of solar disk features (\cite[Chatterjee et al. 2017]{Chatterjee17}), for irradiance reconstruction (\cite[Ermolli et al. 2013]{Ermolli13}) and long-term series (\cite[Chatzistergos et al. 2018]{Chatzistergos18}), are going to be complemented by data obtained with the SAMM (Solar Activity MOF Monitor) telescope. This is a robotic double-channel $2\times$23.5~cm aperture telescope, equipped with magneto\--optical filter (MOF) cells, designed to monitor solar activity with simultaneous measurements of the magnetic field and plasma velocity at two heights in the solar atmosphere. SAMM performs high cadence observations ($< 5$~s) at the Na~I $589.0$~nm and K~I $769.9$~nm lines, with high magnetic sensitivity (10 G). The instrument had first light in Fall 2017. SAMM is a project funded by the Italian Ministry for Economic Development (MiSE) and is a partnership between the Dal Sasso/Avalon Instruments firm and INAF.

\begin{figure}[t]
	\begin{center}
		\includegraphics[scale=0.2825]{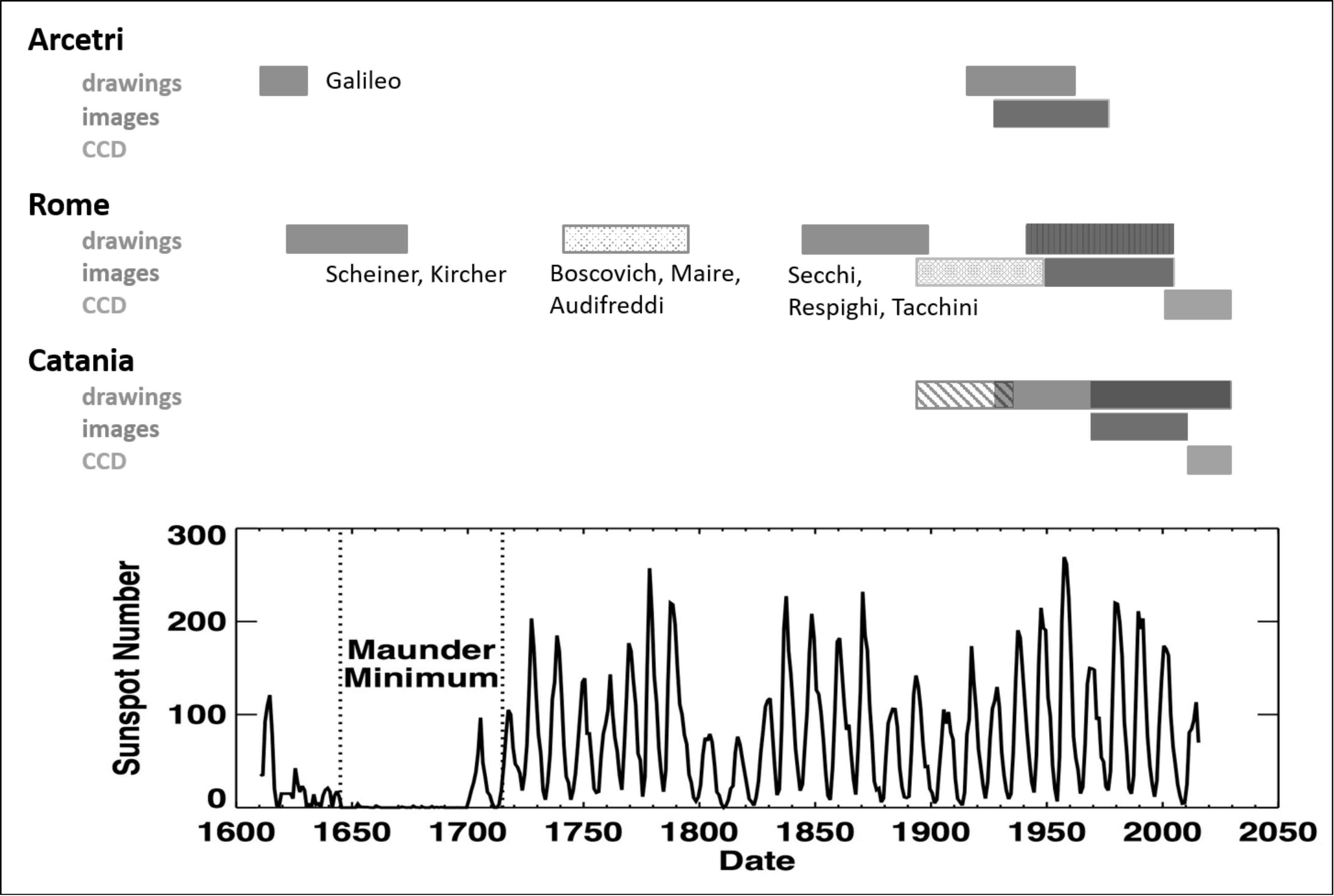} 
		\caption{Synoptic view of the long-term solar data sets in Italy, spanning four centuries of observations. Partially filled bars indicate sparse or not full-disk observations. Shaded areas represent fully digitized data sets.}
		\label{fig1}
	\end{center}
\end{figure}

Daily solar observations at OACT have been taken since 1967 with an Equatorial spar. Today, it includes a 0.15~m Cooke refractor (f=2.3~m), used to make daily drawings of sunspot groups from visual observations, over a 24.5~cm projected image of the Sun; a 0.15~m refractor (f=2.23~m) with a H$\alpha$ Lyot filter for chromospheric and photospheric observations (facility \textsc{OACT:0.15m} for the use in AAS publications); a 0.15~m refractor (f=2.23~m) feeding a H$\alpha$ Halle filter for limb observations of the chromosphere. 

Data products from OACT include full-disk images in the H$\alpha$ line center at $656.28 \pm 0.025$~nm and near the H$\alpha$ continuum at $656.78 \pm 0.025$~nm, with a cadence of 10 minutes and 1 hour, respectively, during the observation time interval (7:30 -- 13:30 CET). The spatial resolution is $2^{\prime\prime}$. Additionally, every observing day a report is compiled according to the USSPS code, based on the sunspot drawing in WL (see \cite[Zuccarello et al. 2011]{Zuccarello11}). 

The OACT observations, which are also used for studying the chromospheric configuration of active regions (e.g., \cite[Guglielmino et al. 2017]{Zuccarello11}, \cite[Romano et al. 2018]{Romano18}), have been recently included as federated products in the Solar Weather Expert Service Centre maintained by the European Space Agency (ESA).

The OACT archive is at the website: \url{http://ssa.oact.inaf.it/oact/index.html}. 

Finally, it has to be recalled that optical observations of the photosphere in recent years were also performed at the Capodimonte Observatory in Naples (1999 --  2010), with the VAMOS (Velocity and Magnetic Observations of the Sun) telescope equipped with MOF cells. This telescope is currently being upgraded to newer components.

\section{Concluding remarks}

A synoptic picture of long-term solar optical data sets acquired in Italy and available to the community is provided in Fig.\,\ref{fig1}.

Regular solar observations performed by several outstanding Italian astronomers since the early $17^{\mathrm{th}}$ century resulted in long series of drawings and images of the photosphere and chromosphere, which now constitute a very valuable collection of historical solar data. An important effort has been undertaken to make the Italian historical observations available in digital form. These series are continued nowadays with observations performed at the Catania and Rome Observatories with modern instruments, which are made available to the community at the websites of the above institutions. 


\end{document}